\documentclass[aps,pra,twocolumn,groupedaddress,amsmath,superscriptaddress]{revtex4-1}
\usepackage{amsmath}
\usepackage{graphicx}
\usepackage{amsfonts}
\usepackage{amssymb}
\usepackage[hidelinks]{hyperref}
\hypersetup{
    colorlinks = true,
    citecolor = blue,
    linkcolor = blue
}
\usepackage{color}
\usepackage{mathrsfs}
\usepackage{isomath}
\usepackage{amsmath}
\usepackage{amsthm}
\usepackage{times,txfonts}

\newcommand{\Tr}{\mbox{Tr}}

\begin{document}

\title{Unconditional security of etanglement-based continuous variable quantum secret sharing}
\author{Ioannis Kogias}
\email{john$_$k$_$423@yahoo.gr}
\affiliation{$\mbox{School of Mathematical Sciences, The University of Nottingham,
University Park, Nottingham NG7 2RD, United Kingdom}$}
\author{Yu Xiang}
\address{State Key Laboratory of Mesoscopic Physics, School of Physics, Peking University, Collaborative Innovation Center of Quantum Matter, Beijing 100871, China}
\author{Qiongyi He}
\address{State Key Laboratory of Mesoscopic Physics, School of Physics, Peking University, Collaborative Innovation Center of Quantum Matter, Beijing 100871, China}
\author{Gerardo Adesso}
\email{gerardo.adesso@nottingham.ac.uk}
\affiliation{$\mbox{School of Mathematical Sciences, The University of Nottingham,
University Park, Nottingham NG7 2RD, United Kingdom}$}

\begin{abstract}
{The need for secrecy and security is essential in  communication. Secret sharing is a conventional protocol to distribute a secret message to a group of parties, who cannot access it individually but need to cooperate in order to decode it. While several variants of this protocol have been investigated, including realizations using quantum systems, the security of quantum secret sharing schemes still remains unproven almost two decades after their original conception. Here we establish an unconditional security proof for continuous variable entanglement-based quantum secret sharing schemes, in the limit of asymptotic keys and for an arbitrary number of players. We tackle the  problem by resorting to the recently developed one-sided device-independent approach to quantum key distribution. We demonstrate theoretically the feasibility of our scheme, which can be implemented by Gaussian states and homodyne measurements, with no  need for ideal single-photon sources or quantum memories. Our results contribute to validating quantum secret sharing as a viable primitive for quantum  technologies.}
\end{abstract}

\date{\today}
\maketitle

\section{Introduction}
\textit{Secret sharing}  \cite{shamir,blakley} is a task where a \textit{dealer} sends a secret $S$ to $n$ (possibly, dishonest) \textit{players}  so that the cooperation of a minimum of $k \leq n$ players is required to decode the secret. 
Protocols that accomplish this are called $(k,n)$-threshold schemes.
The need for such a task appears naturally in many situations, from children's games and online chats, to banking, industry, and military security: the secret message cannot be entrusted to any individual, but coordinated action is required to decrypt it in order to prevent wrongdoings.

For the classical implementation of the simplest $(2,2)$-threshold scheme, Alice, the dealer, encodes her secret into a binary string $S$ and adds to it a random string $R$ of the same length, resulting into the coded cypher $C=S \oplus R$, where ``$\oplus$'' denotes addition modulo $2$. She then sends $R$ and $C$ respectively to the players Bob and Charlie. While the individual parts $R$ and $C$ carry no information about the secret, only by collaboration the players can recover $S$ adding their strings together: $R \oplus C = S$. General $(k,n)$-threshold classical schemes are a bit more involved. Such protocols, however, face the same problem as any other classical key distribution protocol: \textit{eavesdropping}. An eavesdropper, Eve, or even a dishonest player, can intercept the transmission and copy the parts sent from the dealer to the players, thus accessing the secret.

An obvious way to proceed would be for Alice to first employ standard two-party quantum key distribution (QKD) protocols \cite{QKDRMP}, to establish separate secure secret keys with Bob and Charlie, then implement the classical procedure to split the secret $S$ into parts $R$ and $C$, and use the obtained secret keys to securely transmit these parts to each player. The advantage of this protocol, which we call parallel-QKD (pQKD), is that it exploits unconditional security offered by the well-studied two-party QKD against eavesdropping and, very importantly, that it can be   unconditionally secure against any possible dishonest actions of the players. However, pQKD can be demanding in terms of resources, as for a general $(k,n)$ scenario it requires the implementation of $n$ distinct QKD protocols plus the classical procedure to split the secret \cite{shamir}, thus becoming less efficient with increasing $n$. 

An alternative proposal to cope with these difficulties lies in so-called \textit{quantum secret sharing} \cite{Hillery} (QSS) --- alias quantum sharing of a classical secret, distinct from quantum {\it state} sharing \cite{Cleve,Barryno}, in which the secret is a quantum state rather than a classical message ---  which allows for implementing  a $(k,n)$-threshold scheme supported by a \textit{single} classical post-processing, regardless of the number of players $n$.  Unfortunately, as we shall see below, there exists no provably secure QSS scheme at the moment that enjoys the unconditional security of pQKD against both eavesdropping and dishonesty.


Hillery, Bu\v{z}ek, and Berthiaume \cite{Hillery} (HBB, for short) proposed the first (2,2)- and (3,3)-threshold QSS schemes that use multipartite entanglement to split the classical secret, and protect it from eavesdropping and dishonest players in a single go.
Various other entanglement-based (HBB-type) schemes have been proposed \cite{Karlsson, Pan, ZhangMan,ChenLoNoisyGHZ,wu2016CV,Gottesman,MDIQSS,MarinMarkham}, some being more economic in the required multipartite entanglement \cite{EPRpairsDeng,expGisin}, while others allowing for more general $(k,n)$-threshold schemes \cite{NJPLanceLam,Markham,MarkhamQuDit,weedbrook,CVgraphWu}. A different entanglement-based QSS scheme has also been proposed, where entangled states are directly used as secure carriers and splitters of information \cite{bagherinezhad2003quantum}. A few experimental demonstrations have been reported as well \cite{expGisin,expPan,expGaertner,expMarkham}. The security of all current schemes, however, is limited to either plain external eavesdropping unrealistically assuming honest players, or limited types of attacks by eavesdroppers and dishonest participants, yet sharing ideally pure maximally entangled states. Furthermore, all such schemes are vulnerable to participant attack and cheating \cite{Karlsson,CryptanalHBB,DengOpaque}, and no method is currently known to deal with such conspiracies in general, not even in the ideal case of pure shared states.

Zhang, Li, and Man \cite{ZhangLiMan} proposed the first \textit{(n,n)}-threshold scheme that required no entanglement and was claimed to be unconditionally secure.
Although it required perfect single photon sources and quantum memories (rendering it impractical for current technology), it was later shown to be vulnerable to various participant attacks \cite{TrojanHorse,AttackOnImproved}. In the same category of entanglement-free QSS schemes, Schmid \textit{et al.} proposed a protocol based on a single photon \cite{SchmidSingle}; although originally claimed to be unconditionally secure, it was also shown to be vulnerable to participant attacks \cite{AttackOnSchmid,AttackOnImprovedSchmid,KarimiPRA,SingleQudit}. Alternative schemes can be devised to deal with particular attacks  \cite{TrojanHorse,AttackOnImproved,AttackOnSchmid,AttackOnImprovedSchmid}, however there currently exists no rigorous method against arbitrary participant attacks.

To sum up, almost two decades after the conception of QSS, no existing scheme (with or without entanglement) has been proven unconditionally secure against cheating of dishonest players. Hence any practical implementation of secure secret sharing needs to resort to conventional pQKD, while QSS schemes only served so far as a theoretical curiosity.

In this article, we consider a continuous variable  version of an HBB-type scheme. We determine conditions  on the extracted key rate for the secret to be unconditionally secure against both external eavesdropping and arbitrary cheating strategies of dishonest participants, in the limit of asymptotic keys, independently of the shared state, and for arbitrary $(k,n)$-threshold schemes.
The central idea in our approach,  to rigorously deal with arbitrary cheating strategies, is to treat the measurements announced by the players as an input/output of an uncharacterized measuring device (black box), analogously to how (possibly, hacked) measuring devices are treated in device-independent QKD  \cite{DIQKD}.  In practice, this translates into making no assumption about the origin of the players' (possibly, faked) announced measurements, in contrast to previous QSS approaches that considered the players' actions as trusted thus suffering from cheating strategies. The dealer, on the other hand, is regarded as a trusted party with trusted devices, which is a natural assumption for this task. At variance with device-independent QKD, where  the \textit{devices} are untrusted, for the QSS task we treat the \textit{players} themselves as untrusted, independently of their devices. Therefore the framework established in this article, which makes no assumptions about the players' measurements, allows us to prove security against general attacks of eavesdroppers and/or of dishonest players. This is achieved by making a sharp connection with, and extending all the tools of, the recently developed one-sided device-independent QKD (1sDI-QKD) \cite{1sDITomamichel}, in particular for continuous variable systems \cite{walk}, which has been proven unconditionally secure in the limit of asymptotic keys. However, the approach introduced here is general and can be adapted to derive security proofs for discrete variable QSS schemes as well as in the regime of finite keys \cite{CVQKDfinite}.

The paper is organized as follows.
In Section~\ref{Section2} we present our continuous variable QSS protocol, focusing on the $(2,2)$-threshold case. In Section~\ref{Section3} we provide a proof of its unconditional security, adopting techniques from the 1sDI-QKD paradigm. In Section~\ref{Section4} we present extensions to $(k,n)$-threshold schemes and analyze the experimental feasibility of our protocol. In Section~\ref{Section5} we summarize our work and discuss some future perspectives.


\section{The protocol}\label{Section2} For illustration, we first focus on the $(2,2)$-threshold scheme. The trusted dealer Alice prepares a 3-mode continuous variable entangled state, keeps one mode and sends the other modes to the untrusted players, Bob and Charlie, through individual unknown quantum channels.
Alice is assumed to perform homodyne measurements of two canonically conjugate quadratures, $\hat{x}_A = ({\hat{a}+\hat{a}^\dagger})/{\sqrt{2}}$ and $\hat{p}_A=({\hat{a}-\hat{a}^\dagger})/{i\sqrt{2}}$, on her mode, with corresponding outcomes $X_A, P_A$, satisfying $[\hat{x}_A, \hat{p}_A ] =i$ (in natural units with $\hbar = 1$). Bob and Charlie, considered with uncharacterized devices, are entitled to two unspecified  measurements each, labelled by $x_{B(C)}$, $p_{B(C)}$, with corresponding outcomes $X_{B(C)},P_{B(C)}$. Nothing is assumed about the origin of these  measurements.

In our protocol, Alice's goal is to establish a unique secret key, not with Bob's or Charlie's individual measurements (as in standard two-party QKD), but with a collective (non-local) degree of freedom for Bob and Charlie, say $\bar{X}$, that strongly correlates  with one of Alice's quadratures, say $X_A$. The unique secret key can be accessed only when the players communicate their local measurements, i.e., collaborate. For example, if the three parties shared a maximally entangled state and their outcomes were perfectly correlated as $X_A \simeq -X_B + X_C$, one would choose $\bar{X} = -X_B + X_C$ as such collective degree of freedom. 

Alice sends additional copies to Bob and Charlie, and each time all parties randomly choose and measure their parts, getting outcomes $X_i$, $P_i$ respectively, with $i=A,B,C$, until they have a sufficiently long list of correlated data.  Afterwards, all parties announce their measurement choice for each copy and  keep only the data originating from correlated measurements (depending on the shared state). A random subset of this data, chosen by the dealer, is then publicly revealed and used to estimate the size of the secret key (\textit{parameter estimation} step \cite{RennerThesis}) that will provide secure QSS (see below). Finally, if the estimated key is non-zero, Alice proceeds to the conventional classical post-processing steps of \textit{direct reconciliation} and \textit{privacy amplification} \cite{RennerThesis} to create her final secret key, and sends the encrypted secret to Bob and Charlie. However, only when, and if, Bob and Charlie collaborate to form the joint variable $\bar{X}$, can they apply the post-processing instructions on $\bar{X}$ to acquire Alice's secret key. In what follows we will derive conditions on the key rate  to generate secret bits, from the correlations of $X_A$ and $\bar{X}$,  that are unconditionally secure against eavesdropping and dishonest participants.

\section{Security proof}\label{Section3}
\subsection{Security against eavesdropping}
Let us first study security against eavesdropping,  following the QKD work of Walk \textit{et al.} \cite{walk}. Neglecting detector and reconciliation efficiencies, the direct reconciliation  asymptotic secret key rate is known to be lower bounded by the Devetak-Winter formula \cite{devetak},
\begin{equation}\label{devetak}
K \geq I(X_A:\bar{X}) - \chi (X_A:E),
\end{equation}
which finds many uses in quantum information and communication \cite{DIQKD,barbarica,horokarlo,rattomatto,RennerThesis,quantumemory,scaramuc,renner,pirla,fanchinetti,concentrazione,walk}.
Here \begin{equation}
I(X_A:\bar{X}) = H(X_A) - H(X_A|\bar{X})
 \end{equation}
 is the classical mutual information between Alice's variable $X_A$ and the joint variable $\bar{X}$, with $H(X) = - \int dX p(X) \log p(X)$ being the Shannon entropy for a variable $X$ with probability distribution $p(X)$, and
\begin{equation}\label{holevo}
\chi (X_A:E) = S(E) - \int d X_A p(X_A)\, S(\rho^{X_A}_E)
\end{equation}
being the Holevo bound \cite{holevo}, which represents the maximum possible knowledge an eavesdropper can get on the key. The term $S(E) = - \Tr (\rho_E \log \rho_E)$ is the von Neumann entropy of Eve's reduced state $\rho_E$, whereas $\rho^{X_A}_E$ denotes Eve's state conditioned on Alice's measurement of $\hat{x}_A$ with outcome $X_A$. All the logarithms in this paper are taken in base $2$.
All logarithms in this paper are taken in base $2$. A positive value of the right-hand side of \eqref{devetak} implies security of the key against collective attacks of the eavesdropper, and by virtue of Ref.~\cite{renner} also against general coherent attacks. 

Defining the conditional von Neumann entropy
\begin{equation}
S(X_A | E) = H(X_A) +  \int d X_A p(X_A)\, S(\rho^{X_A}_E) - S(E)\,,
 \end{equation}
 and the conditional Shannon entropy
 \begin{equation}H(X_A|X_B) = \int d X_B p(X_B) H(X_A|x_B=X_B)\,,
 \end{equation} with $H(X_A|x_B=X_B) =- \int d X_A p(X_A|X_B) \log p(X_A|X_B)$,
  one can recast the key rate \eqref{devetak} as a balance of conditional entropies,
\begin{equation}\label{devetak2}
K \geq S(X_A|E) - H(X_A|\bar{X}).
\end{equation}
We can now use fundamental {\it entropic uncertainty relations} that provide a lower bound to Eve's uncertainty \cite{PhysRevLett.103.020402,quantumemory, furrer, franklieb, entureview,ferenczi},
\begin{equation}\label{uncertainty}
 S(X_A|E) + S(P_A|BC)  \geq \log 2 \pi,
\end{equation}
for the derivation of which Alice's canonical commutation relations have been assumed, while Eve is assumed to purify the state shared by Alice, Bob and Charlie, i.e., $\rho_{ABC} = \Tr_E \left( |\Psi_{ABCE}\rangle \langle \Psi_{ABCE}|  \right)$.
Substituting the uncertainty relation \eqref{uncertainty} back into \eqref{devetak2} and recalling that $S(P_A|BC) \leq S(P_A|\bar{P}) = H(P_A|\bar{P})$ (since measurements cannot decrease the entropy), where $\bar{P}$ is a joint variable for Bob and Charlie optimally correlated with Alice's momentum $P_A$, we get
\begin{equation}
K \geq \log 2 \pi - H(X_A|\bar{X}) - H(P_A|\bar{P})\,,
 \end{equation} i.e., a bound on the key rate (hence, on Eve's maximal knowledge on the key $X_A$) only involving conditional Shannon entropies, that can be estimated using the announced measurement outcomes during the parameter estimation stage.

 To make the bound even more accessible, we proceed to express it only in terms of second moments, instead of dealing with conditional probability distributions.
For this aim, we recall that the Shannon entropy of an arbitrary probability distribution is maximized by a Gaussian distribution of the same variance. In other words, $H(X_A |\bar{X}) \leq H_G (X_A |\bar{X}) = \log \sqrt{ 2 \pi e V_{X_A |\bar{X}}}$, where
\begin{equation}
V_{X_A |\bar{X}} = \int  d\bar{X} p(\bar{X}) \left( \langle X_A^2 \rangle_{\bar{X}} - \langle X_A \rangle^2_{\bar{X}} \right)
 \end{equation}
 is the minimum inference variance of Alice's position outcome when the joint outcome $\bar{X}$ is known; and similarly for $H(P_A|P)$. The final key rate is then bounded as follows,
\begin{equation}\label{devetakfinal}
K \geq - \log \left(   {e \sqrt{V_{X_A |\bar{X}}V_{P_A |\bar{P}}}}   \right).
\end{equation}
We see that a nonzero key rate (secure against eavesdropping) can be achieved when $E_{A|BC} \equiv V_{X_A |\bar{X}}V_{P_A |\bar{P}} < e^{-2} $.

\subsection{Security against dishonesty}
We derived conditions such that Alice's key is secure from eavesdropping and the players can safely obtain the key whenever they decide to collaborate. However, one needs to consider also the potential cheating strategies of the players themselves.

Suppose now that Bob is a dishonest player. His goal would be to guess Alice's key (hence, access the secret) using solely his own local measurements $x_B, p_B$, entirely bypassing the required collaboration with Charlie. A most general cheating strategy for Bob would be: first, to secretly intercept Charlie's mode during its transmission using general coherent attacks to increase his knowledge on Alice's key; and second, to lie about his measurements. A positive key rate in \eqref{devetakfinal} does not guarantee security against such general participant attacks and cheating.

Here we  derive additional conditions on the key rate so that Bob cannot cheat or access the secret by himself. Our central observation is to reconsider the Devetak-Winter formula \eqref{devetak} and treat now Bob as an eavesdropper, together with Eve. This means that in the Holevo bound $\chi (X_A:E)$ defined in (\ref{holevo}), that expresses the knowledge of party $E$ on the key $X_A$, we will include Bob himself. This leads to a modified Devetak-Winter formula,
\begin{equation}\label{devetak3}
K \geq I(X_A:\bar{X}) - \chi (X_A:EB),
\end{equation}
where $EB$ refers now to the unknown joint quantum state of Eve and Bob. A positive key rate in \eqref{devetak3} would imply security of Alice's key against joint general attacks by Bob and Eve on Charlie's system. Also, Bob and Eve's maximum knowledge of the key, $\chi (X_A:EB)$, can be upper bounded as seen below using Alice and Charlie's measurements, independently of Bob's (possibly, faked) announced measurements, therefore
providing security against Bob's cheating.  The uncertainty relation that we will use to bound Bob and Eve's knowledge will be a slightly modified version of \eqref{uncertainty},
\begin{equation}\label{uncertaintyEB}
 S(X_A|EB) + S(P_A|C)  \geq \log 2 \pi.
\end{equation}
Following similar steps as previously described, we thus end up with the following novel bound on the key rate,
\begin{equation}\label{devetakfinalEB}
K \geq -\log \left(  {e \sqrt{V_{X_A |\bar{X}}V_{P_A |P_C}}}   \right).
\end{equation}
Notice that the key rate bound in \eqref{devetakfinalEB} is smaller than the  one in \eqref{devetakfinal} that did not take dishonesty into account, due to $V_{P_A |\bar{P}} \leq V_{P_A |P_C}$, which is expected since the eavesdroppers' knowledge on the key is increased by including Bob together with Eve.

To intuitively understand why this condition prohibits any cheating from Bob, we recall first that the key is generated solely by the $X_A, \bar{X}$ outcomes. By examination of the uncertainty relation \eqref{uncertaintyEB}, taking into account that $\log \sqrt{ 2 \pi e V_{P_A |P_C}} \geq S(P_A|C)$, we see that the better Charlie can estimate Alice's momentum (i.e., the smaller $S(P_A|C)$) the larger Bob and Eve's ignorance should be on the key elements $X_A$. The previous condition \eqref{devetakfinal}, not accounting for participant dishonesty, only demanded that  $ S(P_A |BC)$ is small enough, which can be true even if $S({P_A |C})$ is arbitrarily large, thus allowing Bob to reach good knowledge of the key (i.e, small $S({X_A |EB})$), through \eqref{uncertaintyEB}.


We can also account for Charlie's dishonesty in an exactly analogous manner (just replace $B \leftrightarrow C$ above), leading us to
\begin{equation} \label{devetakfinalEC}
K \geq -\log \left( {e \sqrt{V_{X_A |\bar{X}}V_{P_A |P_B}}}   \right).
\end{equation}

Putting everything together, the final bound on the asymptotic key rate to provide unconditional security against general attacks of an eavesdropper, and against arbitrary (individual) cheating methods of both Bob and Charlie, which include the announcement of faked measurements and general attacks of Bob on Charlie's system and of Charlie on Bob's system, is:
\begin{equation}\label{devetakfinalfinal}
\begin{split}
K & \geq I(X_A:\bar{X}) - \max \{\chi (X_A:EB),\,\, \chi (X_A:EC) \} \\
  & \geq   -\log \left( e \sqrt{ V_{X_A |\bar{X}} \cdot  \max \{ V_{P_A |P_C}, V_{P_A |P_B}   \} }   \right),
\end{split}
\end{equation}
which is  the minimum of the bounds \eqref{devetakfinalEB} and \eqref{devetakfinalEC}.
A positive key rate \eqref{devetakfinalfinal} remarkably provides security against all kinds of attacks that existing QSS protocols suffered from (e.g., fake announced measurements \cite{Karlsson}, Trojan horse attacks \cite{TrojanHorse}, etc.), for the sole reason that the players Bob and Charlie are not assumed to be performing trusted quantum operations but are treated as black boxes, in contrast to all previous schemes. 

\begin{figure}[t]
\includegraphics[width=8.5cm]{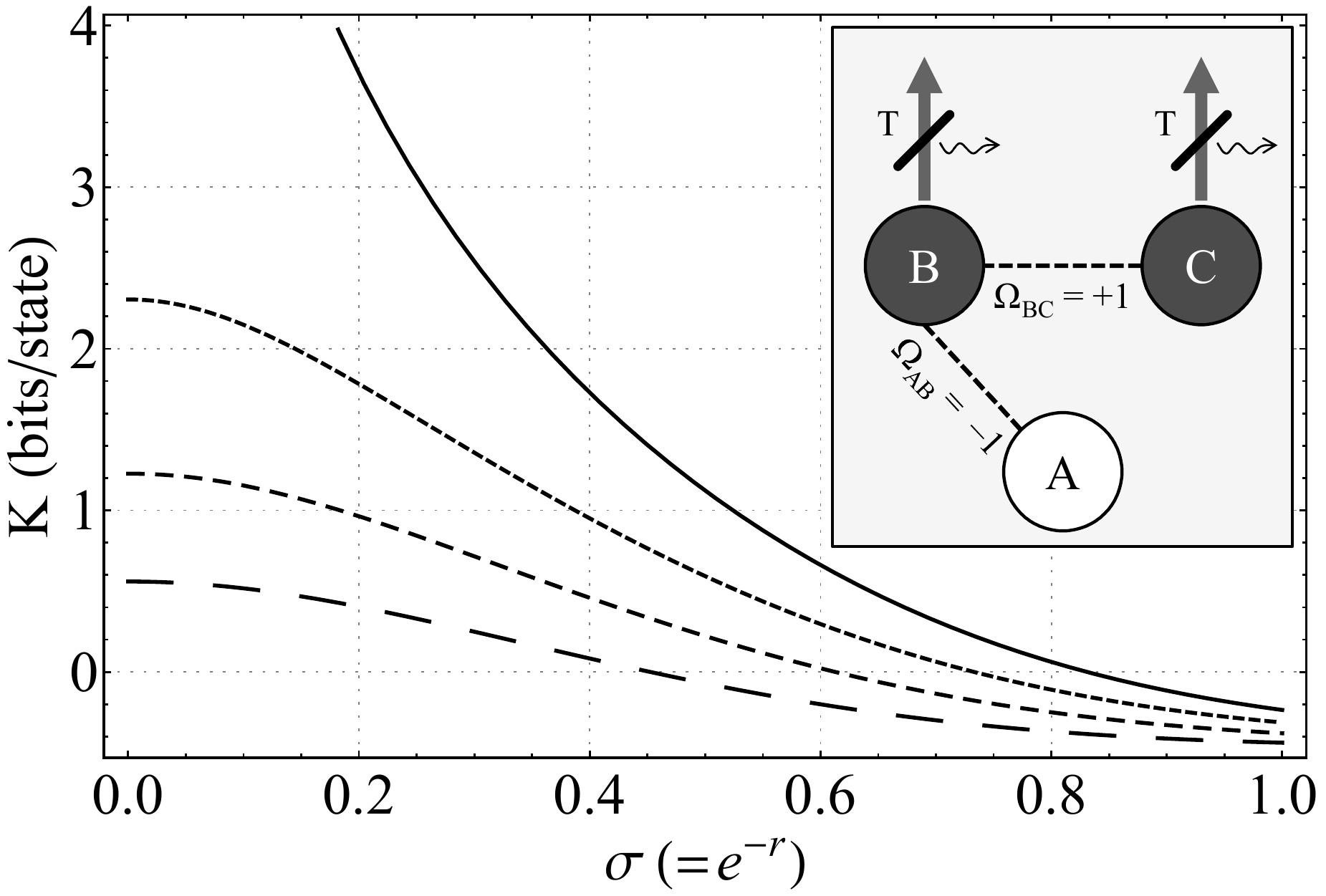}
\caption{The QSS  secure key rate $K$, Eq.~\eqref{devetakfinalfinal}, is plotted against the squeezing $r$ of a 3-mode noisy Gaussian cluster state, obtained from a pure state \cite{weedbrook} $\hat{U}_{AB} \hat{U}_{BC} |r\rangle_A |r\rangle_B |r\rangle_C$, with $\hat{U}_{ij} = \exp \left( \Omega_{ij} \hat{x}_i \hat{x}_j \right)$, after Bob and Charlie's modes undergo individual pure-loss channels (i.e., quantum-limited attenuating channels), each modelled by a beam splitter with transmissivity $T$  and zero excess noise (see inset). From top to bottom, the curves correspond to  $T=1,\,\,0.95,\,\,0.9,\,\,0.85$. All parties are assumed to be performing homodyne measurements of $\hat{x}_i$,$\hat{p}_i$, with $i=A,B,C$.
The current experimentally accessible squeezing is limited to $r \lesssim 1.15$ (10dB), or $\sigma \gtrsim 0.32$ \cite{SchnabelPRL,SchnabelOptExpr}, in which regime a nonzero $K$ is still guaranteed for sufficiently large $T$, demonstrating the feasibility of our scheme.}
\label{Keyrates}
\end{figure}

\section{Discussion and extensions}\label{Section4}
In Fig.~\ref{Keyrates} we demonstrate the feasibility of the protocol in a concrete realization,  where the key rate \eqref{devetakfinalfinal} is plotted against the squeezing degree of a noisy tripartite entangled Gaussian cluster state. Notice that the same key rate can also be achieved by an equivalent protocol that solely requires bipartite entanglement (that would represent the so-called prepare-and-measure counterpart to the presented protocol, borrowing a QKD terminology), thus further reducing the technological requirements for the state preparation. More generally, given the recent progress in the generation of large-scale continuous variable entangled states \cite{shanxi8,armstrong2015multipartite,TrepsNature,TrepsPRL,Pfister,Furusawa10k,Furusawainf}, we expect our secure protocol to be recognized as a competitive candidate for practical QSS, in alternative to conventional pQKD.

Finally, we show how to generalize the secret key rate bound \eqref{devetakfinalfinal} to any $(k,n$)-threshold QSS scheme. To start with,  let us denote the $n$ players as $B_1, B_2,\ldots,B_n$. A $(k,n)$-threshold scheme has two requirements:
first, no collaboration of any $k-1$ players should be able to access the secret. We incorporate this requirement into Eq.~\eqref{devetakfinalfinal} by considering all possible combinations of $k-1$ out of $n$ players, the total number of which equals the binomial coefficient ${{n}\choose{k-1}}$, as potential collaborative eavesdroppers, and choosing the maximum Holevo information over all collaborations to attain the maximum possible knowledge on the key by any of these groups.
Second, any collaboration of $k$ players, known as the {\it access structure}, should be able to decode the message.
Let us attribute a joint variable $\bar{X}_i$ to each $k$-player collaboration correlated to Alice's $X_A$, with $i=1,\ldots,{{n}\choose{k}}$. This amounts
to Alice sending as much error-correction information as needed, such that even the $k$-player collaboration least correlated to Alice, i.e., with the smallest $I(X_A:\bar{X}_i)$, can access her key. Taking the above into account, the key rate of the protocol will be,
\begin{equation}\label{devetakfinalfinalgeneral}
\begin{split}
K  \geq \min \big\{ & I(X_A:\bar{X}_1),\ldots, I(X_A:\bar{X}_{{{n}\choose{k}}})   \big\}    \\
& - \max \big\{\chi (X_A:ES_1),\ldots, \chi (X_A:ES_{{{n}\choose{k-1}}}) \big\},
\end{split}
\end{equation}
where $S_i$ denotes a particular sequence of $k-1$ players, e.g., $S_1 = B_1\cdots B_{k-1}$.
A positive value of the right-hand side of Eq.~\eqref{devetakfinalfinalgeneral} guarantees  unconditional security of our QSS protocol against eavesdropping and arbitrary collaborative cheating strategies of any group of $k-1$ potentially dishonest players. This analysis readily extends to arbitrary access structures, where a subset of privileged players can access the key.


\section{Conclusions}\label{Section5} We presented a feasible entanglement-based continuous variable
QSS scheme, and derived sufficient conditions for the protocol's secret key rate to provide, for the first time, unconditional security of the dealer's classical secret against general attacks of an eavesdropper and arbitrary cheating strategies, conspiracies and attacks of the (possibly, dishonest) players, for all $(k,n)$-threshold schemes, and in the limit of asymptotic keys.

 In our  approach, we crucially identified  the most physically relevant framework for QSS to be the 1sDI setting, treating the dealer as a trusted party with characterized devices and the players' devices as black boxes. The natural separation of roles between dealer and players renders QSS a well-suited task for the 1sDI setting, even more than two-party QKD itself \cite{oneSidedDIQKD}. At the same time, this observation enables us to adopt and generalize conventional 1sDI-QKD techniques to establish security of entanglement-based QSS, as demonstrated in this paper.
 Incidentally, while the resource behind 1sDI-QKD is known to be (bipartite) {\em steering} \cite{wiseman}, a quantum correlation stronger than plain entanglement \cite{ent} and weaker than Bell-nonlocality \cite{nonloc}, one could suspect a similar connection in the present multiuser scenario. In a companion paper \cite{Yucompanion}, we show in fact that \textit{multipartite} steering \cite{HeReidGenuine,armstrong2015multipartite} empowers secure QSS, providing an operational interpretation for a genuine multipartite continuous variable steering measure.

Our work opens many avenues for further exploration. The presented security proof rests on general principles and can be extended from asymptotic to finite keys \cite{CVQKDfinite}, suitable for concrete applications, and also to discrete variable systems, used in the original QSS definition \cite{Hillery}; this will be the subject of future work. Moreover, although we provided  sufficient security conditions for all $(k,n)$-threshold schemes, the identification of optimal families of states maximizing the key rate for each scheme was left open and will be addressed elsewhere.

Finally, our results pave the way for an unconditionally secure experimental demonstration of QSS, enabling its use in next-generation quantum communication networks.

\begin{acknowledgments}
I.K. and G.A. acknowledge funding from the European Research Council under Grant No.~637352 (ERC StG GQCOP); Q.H. acknowledges the support of the National Natural Science Foundation of China under Grants No.~11274025 and No.~61475006. We thank A. Winter, N. Walk, E. Woodhead and, particularly, A. Leverrier, for fruitful discussions. I. K. thanks R. Hawkins for proofreading the manuscript.
\end{acknowledgments}


%

\end{document}